# THE CHERNOFF LOWER BOUND FOR SYMMETRIC QUANTUM HYPOTHESIS TESTING

By Michael Nussbaum[1] and Arleta Szkoła[2]

*Cornell University and Max Planck Society*

We consider symmetric hypothesis testing in quantum statistics, where the hypotheses are density operators on a finite-dimensional complex Hilbert space, representing states of a finite quantum system. We prove a lower bound on the asymptotic rate exponents of Bayesian error probabilities. The bound represents a quantum extension of the Chernoff bound, which gives the best asymptotically achievable error exponent in classical discrimination between two probability measures on a finite set. In our framework, the classical result is reproduced if the two hypothetic density operators commute.

Recently, it has been shown elsewhere [*Phys. Rev. Lett.* **98** (2007) 160504] that the lower bound is achievable also in the generic quantum (noncommutative) case. This implies that our result is one part of the definitive quantum Chernoff bound.

**1. Introduction.** One typical problem in hypothesis testing is to decide between two equiprobable hypotheses, say $H_0$ and $H_1$, where $H_i$ assumes that the observed data are generated by an i.i.d. process with law $P_i$, $i = 0, 1$. In the classical setting, $P_0, P_1$ are probability measures on a measurable space, the sample space. One discriminates between them by means of test functions, which are nonnegative measurable functions on the $n$-fold product sample space. An error occurs if, according to the given decision rule based on the value of the test function, one accepts hypothesis $H_0$ while the data are generated with law $P_1$, or vice versa.

If one declares one of the hypotheses to be the null hypothesis and the other one the alternative, then errors occurring while the null hypothesis is

Received April 2007; revised December 2007.
[1]Supported in part by NSF Grant DMS-03-06497.
[2]Supported in part by German Research Foundation (DFG) via the project "Entropy, Geometry and Coding of Large Quantum Information Systems."

true are called "of first kind," otherwise "of second kind." Due to Stein's lemma there exists test functions maintaining a given upper bound $\alpha$ on the error probability of first kind, such that the probability of error of the second kind decreases to 0 with the optimal asymptotic rate exponent equal to the Kullback–Leibler distance from the null hypothesis to the alternative. Sanov's theorem extends this result to the case where, instead of a single measure $P_0$, a family $\Omega$ of measures is associated with the null hypothesis. Then, the negative Kullback–Leibler distance from the set $\Omega$ to $P_1$ gives the minimal asymptotic error exponent ([19], see also [7]).

In *symmetric* hypothesis testing one treats the errors of first and second kind in a symmetric way. We will focus here on the *Bayesian error probability*, which is the average of the two kinds of error probabilities. It is minimized by the likelihood ratio test and vanishes exponentially fast as the sample size $n$ tends to infinity. The corresponding optimal asymptotic rate exponent is equal to the *Chernoff bound*

$$\tag{1} \inf_{0 \leq s \leq 1} \log \int p_0^{1-s}(\omega) p_1^s(\omega) \mu(d\omega)$$

pertaining to probability measures $P_0$ and $P_1$, with respective densities $p_0$ and $p_1$ (wrt dominating measure $\mu = P_0 + P_1$). These results go back to papers by Chernoff and Hoeffding [6, 12]. Chentsov and Morozova [5] present a thorough and illuminating discussion of the Chernoff bound, relating it to the differential geometry of statistical inference.

If the data are obtained from quantum systems, then one has to replace probability measures by quantum states, that is, by normalized positive linear functionals on an appropriate algebra of observables. In the present paper, this is assumed to be the algebra of linear operators on a finite-dimensional complex Hilbert space. One discriminates between two states $\rho_0$ and $\rho_1$ by means of quantum tests, which are defined as positive operator valued measures on $n$-fold tensor products of the algebra of observables of a single quantum system. Here, we employed the standard language of quantum mechanics; throughout the paper, however, we will utilize an elementary and accessible mathematical framework based on complex linear algebra only. It will become apparent that quantum tests are analogs of test functions defined on finite sample spaces and their $n$-fold products.

While the basic problems in nonsymmetric quantum hypothesis testing (pertaining to $\alpha$-tests) were solved in [11, 18] and [3] by obtaining quantum versions of Stein's lemma and Sanov's theorem, the case of discrimination (or equally weighted hypotheses) has not yet received full treatment. Although quantum tests minimizing the generalized Bayesian error probabilities were constructed about 30 years ago by Helstrom and Holevo [10, 13], a closed form expression for the optimal asymptotic quantum error exponent similar to the classical Chernoff distance remained an open problem. A reason is that there is no obvious canonical way to extend (1) to a quantum setting. On the very formal level, due to noncommutativity effects, there are different nonequivalent ways of generalizing the distance. In [18], Ogawa and Hayashi list three candidates for the optimal quantum rate exponent, relying on three different extensions of the target function in the variational formula (1). However, two of these candidate expressions are not well defined if the hypotheses are not faithful states, that is, if the associated density operators do not have full rank.

Recently, the problem of symmetric quantum testing was treated by Kargin [14], with partial progress toward the definitive Chernoff bound. Lower and upper bounds on the optimal error exponent in terms of fidelity between the two density operators were given; the lower bound was shown to be sharp in the case that one of the density operators has rank one (i.e., represents a pure quantum state). We remark that fidelity is a notion of distinguishability between density operators which is frequently used in quantum information theory (see, e.g., [8, 16]).

Our main result, which we formulate rigorously in Section 2, states that $\inf_{0 \leq s \leq 1} \log \text{Tr}[\rho_0^{1-s} \rho_1^s]$ is a lower bound on the general asymptotic error exponent, $\rho_0$ and $\rho_1$ being density operators replacing the probability densities $p_0$ and $p_1$ of the classical setting. We remark that our quantum bound coincides with one of the three candidates for a quantum Chernoff bound discussed in [18]. We prove the main theorem in Section 3. Recently, Audenaert et al. have shown in [1] that in accordance with our conjecture stated in a previous version of the present work, [17], the lower bound is indeed achievable. This justifies referring to it as the quantum Chernoff bound.

**2. Mathematical setting and the main theorem.** For an elementary introduction to quantum statistics with physical background, see Gill [9]. We will describe here only the formalism for the simplest possible nonclassical setup of discrimination between two hypotheses. A *density matrix* $\rho$ is a complex, self-adjoint, positive $d \times d$ matrix satisfying the normalization condition $\text{Tr}[\rho] = 1$, where $\text{Tr}[\cdot]$ is the trace operation. Here "positive" means nonnegative definite. We identify a density matrix with a state of a quantum system; we also use "matrix" and "operator" interchangeably. The two hypotheses are described by two states, $H_0 : \rho = \rho_0$ and $H_1 : \rho = \rho_1$.

Physically discriminating between them corresponds to performing a measurement on the quantum system. Mathematically a measurement with $k$ possible outcomes is associated to a set of positive $d \times d$ matrices $\{r_1, \ldots, r_k\}$ adding up to the unit matrix. When the state is $\rho$ then the probability of the $i$th outcome is $\text{Tr}[\rho r_i]$. In analogy to classical hypothesis testing one accepts $H_0$ or $H_1$ according to a decision rule based on the outcome of a measurement. In this case, there are $k = 2$ possible outcomes and any appropriate measurement may be written $\{\mathbf{1} - r, r\}$, where $r$ is a complex self-adjoint positive matrix satisfying the inequality $0 \leq r \leq \mathbf{1}$. Here, $\mathbf{1}$ is the unit matrix and $\leq$ is in the sense of matrix order, that is, $\mathbf{1} - r$ is positive (nonnegative definite). We will mostly make reference to this measurement by its $r$ element, the one corresponding to the alternative hypothesis. Then, $\text{Tr}[\rho r]$ is the overall probability of rejecting $H_0$ when $\rho$ is the true state. Accordingly, $\text{Tr}[\rho_0 r]$ is the *error probability of first kind* and $\text{Tr}[(\mathbf{1} - r)\rho_1] = 1 - \text{Tr}[\rho_1 r]$ is the *error probability of second kind*. When both $\rho_0$, $\rho_1$ and also $r$ are diagonal matrices, then the setup reduces to the classical testing problem for two probability measures on an appropriate index set $\Omega$, $|\Omega| = d$, given by $\rho_0$, $\rho_1$, respectively. The same is true when $\rho_0$, $\rho_1$ have the same set of eigenvectors; then, $\rho_0$, $\rho_1$ are said to commute (*commutative case*). In this sense, commuting states describe the classical discrimination problem between two probability measures on a finite sample space $\Omega$, as a special case of the present quantum setting.

A *pure state* is given by a density matrix which has rank 1, which means it is a projection onto a subspace of (complex) dimension one. We will also use the following notation: we set $\mathcal{H} = \mathbb{C}^d$, with the understanding that $\mathcal{H}$ can be any $d$-dimensional complex Hilbert space, and we write $\mathcal{B}(\mathcal{H})$, $\mathcal{B}(\mathcal{H}^{\otimes n})$ for the set of complex $d \times d$ or $d^n \times d^n$ matrices, respectively. In the bra-ket notation, $|v\rangle$ and $\langle v|$ denote a vector in $\mathcal{H}$ and its dual vector with respect to the scalar product in $\mathcal{H}$ (essentially a column and a row vector). A one-dimensional projection onto a subspace of $\mathcal{H}$, spanned by a unit vector $v$, may be written as $|v\rangle\langle v|$. It is a density operator of a pure state.

The above describes the basic setup where the finite dimension $d$ is arbitrary. We consider the quantum analog of having $n$ i.i.d. observations. For this, the two hypotheses are assumed to be $\rho_0^{\otimes n}$ and $\rho_1^{\otimes n}$ for two basic $d$-dimensional states $\rho_0$, $\rho_1$, where $\rho^{\otimes n}$ is the $n$-fold tensor product of $\rho$ with itself. (Recall that the tensor product $a \otimes b$ of two matrices is a matrix which consists of blocks $a_{ij}b$, arranged according to the indices $i, j$. Thus, $\rho_0^{\otimes n}$ is a $d^n \times d^n$ matrix.) The tests $r_n$ now operate on the states $\rho_0^{\otimes n}$ and $\rho_1^{\otimes n}$, that is, their dimension is $d^n \times d^n$, but they need not have tensor product structure. The corresponding *Bayesian error probability* is

$$\text{Err}(r_n) := \tfrac{1}{2}\text{Tr}[(r_n \rho_0^{\otimes n} + (\mathbf{1} - r_n)\rho_1^{\otimes n})]$$
$$= \tfrac{1}{2}(1 - \text{Tr}[r_n(\rho_1^{\otimes n} - \rho_0^{\otimes n})]).$$

The optimal hypothesis tests minimizing the error probability are known to be the *Holevo–Helstrom hypothesis tests* [10, 13]. They are given for each $n \in \mathbb{N}$ by the projections

$$\Pi_n^* := \text{supp}(\rho_1^{\otimes n} - \rho_0^{\otimes n})_+,$$

where supp $a$ denotes the support projection of a linear operator $a$ and $a_+$ means the positive part of a self-adjoint operator $a$. Thus, if $a = \sum_i \lambda_i E_i$ is the spectral decomposition using projections $E_i$, then $a_+ := \sum_{\lambda_i > 0} \lambda_i E_i$ and $\operatorname{supp} a_+ = \sum_{\lambda_i > 0} E_i$. Indeed, we have for an arbitrary test operator in $\mathcal{B}(\mathcal{H}^{\otimes n})$

$$\begin{aligned}
\operatorname{Err}(r_n) &= \tfrac{1}{2}(1 - \operatorname{Tr}[r_n(\rho_1^{\otimes n} - \rho_0^{\otimes n})]) \\
&\geq \tfrac{1}{2}(1 - \sup\{\operatorname{Tr}[\tilde{r}(\rho_1^{\otimes n} - \rho_0^{\otimes n})] \colon \tilde{r} \in \mathcal{B}(\mathcal{H}^{\otimes n}) \text{ test}\}) \\
&= \tfrac{1}{2}(1 - \sup\{\operatorname{Tr}[\Pi(\rho_1^{\otimes n} - \rho_0^{\otimes n})] \colon \Pi \in \mathcal{B}(\mathcal{H}^{\otimes n}) \text{ projection}\}) \\
&= \tfrac{1}{2}(1 - \operatorname{Tr}[\Pi_n^*(\rho_1^{\otimes n} - \rho_0^{\otimes n})]) = \tfrac{1}{2}(1 - \tfrac{1}{2}\|\rho_1^{\otimes n} - \rho_0^{\otimes n}\|_1),
\end{aligned}$$

where $\|a\|_1 = \operatorname{Tr}[a_+] + \operatorname{Tr}[a_+ - a]$ is the generalization of the $L_1$-norm. Note that the last line above gives an exact closed form expression of the best error probability for every $n$, but its asymptotics as $n \to \infty$ (rate of exponential decay) is the subject of the present paper.

The Holevo–Helstrom tests $\Pi_n^*$ are noncommutative generalizations of the likelihood ratio tests: if the hypotheses $H_0$ and $H_1$ correspond to commuting density operators $\rho_0$ and $\rho_1$, then, for all $n \in \mathbb{N}$, the Holevo–Helstrom projections $\Pi_n^*$ commute with $\rho_0^{\otimes n}$ and $\rho_1^{\otimes n}$, also. The density operators $\rho_i$ may be completely specified by their eigenvalues forming discrete probability measures $P_i$, $i = 0, 1$, on an appropriate index set $\Omega$, $|\Omega| = d$ for the mutually commuting spectral projectors on $\mathcal{H}$. For each $n \in \mathbb{N}$, the set of eigenvalues of the tensor product $\rho_i^{\otimes n}$, $i = 0, 1$, corresponds to the respective product measure $P_i^n := \prod_{j=1}^n P_i$ on the Cartesian product $\Omega^n := \times_{i=1}^n \Omega$ while the Holevo–Helstrom projection $\Pi_n^*$ generalizes the indicator function $\lambda_n^* = \mathbf{1}\{p_1^n - p_0^n > 0\}$ on $\Omega^n$. Here, $p_i^n$ denote the probability densities of the product measures $P_i^n$. We note that $\lambda_n^*$ is the well-known maximum likelihood decision. It takes the value 1, which corresponds to a decision in favor of $H_1$, on samples $x \in \Omega^n$ for which the density value (or likelihood) $p_1^n(x)$ is larger than $p_0^n(x)$.

The classical Bayesian error probability $\operatorname{Err}(\lambda)$, of a test function $\lambda$ ($0 \leq \lambda \leq 1$), is defined by

$$(2) \qquad \operatorname{Err}(\lambda) := \tfrac{1}{2}(E_{P_0}\lambda + E_{P_1}(1 - \lambda))$$

where $E_P$ stands for expectation under the law $P$. The quantity $\operatorname{Err}(\lambda)$ averages over both possible sources of error with equal weights $1/2$. In the more general situation, the weights are specified by the a priori probabilities $(\pi_0, \pi_1)$ for $H_0$ or $H_1$ to occur, that is, $\operatorname{Err}(\lambda) := \pi_0 E_{P_0}\lambda + \pi_1 E_{P_1}(1 - \lambda)$.

As already mentioned in the [Introduction](#), the Bayesian error probability $\operatorname{Err}(\lambda_n^*)$ vanishes, as $n \to \infty$, with a minimal asymptotic rate exponent equal to the *Chernoff bound* $\delta(P_0, P_1)$:

$$(3) \qquad \lim_{n \to \infty} \frac{1}{n} \log \operatorname{Err}(\lambda_n^*) = \delta(P_0, P_1) := \inf_{0 \leq s \leq 1} \log \sum_{x \in \Omega} p_0^{1-s}(x) p_1^s(x).$$

We remark that

(4) $$\sum_{x\in\Omega} p_0^{1-s}(x)p_1^s(x) =: A(s), \qquad s \in [0,1],$$

represent the normalization factors of the parametric family of probability measures

$$p_s(x) := \frac{1}{A(s)} p_0^{1-s}(x)p_1^s(x), \qquad x \in \Omega.$$

The family is called a *Hellinger arc* in the literature. It interpolates between $p_0$ and $p_1$ if their supports $D_0, D_1 \subseteq \Omega$ coincide. Otherwise, $p_s$, $s \in [0,1]$, is discontinuous (in the Euclidian metric of $\mathbb{R}^{|\Omega|}$) at the endpoints $s = 0, 1$ such that over the open parameter interval $(0,1)$ it represents an interpolation between the densities of the conditional probabilities $Q_0 := P_0(\cdot|B)$ and $Q_1 := P_1(\cdot|B)$, where $B := D_0 \cap D_1$.

There is an equivalent expression for the Chernoff bound (3) in terms of the KL-distance (relative entropy):

(5) $\delta(P_0, P_1) = \inf_{s \in [0,1]} (-(1-s)K(Q_s\|Q_0) - sK(Q_s\|Q_1) + \log \pi_0^{1-s}\pi_1^s),$

where $Q_s$ denotes the conditional probability $P_s(\cdot|B)$, for $s \in [0,1]$, and $\pi_i := P_i(B)$, for $i = 0, 1$. Observe that if the supports $D_0$ and $D_1$ coincide, that is, $B = \Omega$, then the target function in (5)—we will refer to it as $H(s)$ in the sequel—becomes simply $-(1-s)K(P_s\|P_0) - sK(P_s\|P_1)$. What is remarkable is that in this case we have

$$\delta(P_0, P_1) = -K(P_\sigma\|P_0) = -K(P_\sigma\|P_1),$$

where the parameter $\sigma \in [0,1]$ is uniquely defined by the second equality above. In the generic case of possibly different supports, a modified version of the above formula is valid. One distinguishes two cases: if there exists a $\sigma \in (0,1)$ such that $H'(\sigma) = 0$, which is equivalent to $K(Q_\sigma\|Q_0) - K(Q_\sigma\|Q_1) = \log(\pi_0/\pi_1)$, then

$$\delta(P_0, P_1) = -K(Q_\sigma\|P_0) + \log \pi_0 = -K(Q_\sigma\|P_1) + \log \pi_1.$$

Otherwise, the infimum in (5) is attained either at $s = 0$ or at $s = 1$, and the corresponding values of the Chernoff bound are $\log \pi_0$ and $\log \pi_1$.

The identity (5) and the other claims in the above paragraph follow from (23) in the Appendix and attendant reasoning. To our knowledge, no quantum generalization of (5) has yet been found.

In the following theorem we formulate the classical result (3) for the general case of probability measures $P_0, P_1$ on an arbitrary measurable space $(\Omega, \Sigma)$, not necessarily finite. Consider the Bayesian error probability of discrimination between $P_0, P_1$ by means of test functions $0 \leq \lambda \leq 1$:

(6) $$\Delta(P_0, P_1) := \inf_{\lambda \text{ test function}} \mathrm{Err}(\lambda)$$

where $\text{Err}(\lambda)$ is given by (2). Let $\lambda^*$ be the maximum likelihood test function $\lambda^* = \mathbf{1}\{p_1 - p_0 > 0\}$ on $\Omega$, in terms of densities $p_0, p_1$, for some dominating measure $\mu$. It is well known that $\Delta(P_0, P_1)$ can be expressed as

$$(7) \qquad \Delta(P_0, P_1) = \text{Err}(\lambda^*) = \tfrac{1}{2} \int \min(p_0, p_1) \, d\mu.$$

THEOREM 2.1. *Let $P_0, P_1$ be two probability measures on $(\Omega, \Sigma)$. For product measures $P_0^n$, $P_1^n$ corresponding to $n$ i.i.d. observations $\omega_1, \ldots, \omega_n$, all having law $P_0$ or $P_1$, the Bayesian error probability satisfies*

$$(8) \qquad \lim_{n \to \infty} n^{-1} \log \Delta(P_0^n, P_1^n) = \inf_{0 \le s \le 1} \log \int p_1^s p_0^{1-s} \, d\mu,$$

*where $p_i = dP_i/d\mu$, $i = 0, 1$, $\mu := P_0 + P_1$.*

For strictly positive $p_0$ and $p_1$ with $p_0 \ne p_1$, the proof can be found in the literature (cf., e.g., [5], page 164, or for finite sample space [7], page 312). For completeness, we present a proof for the general case of possibly different support of $P_0, P_1$ in the Appendix. Indeed, if $P_0, P_1$ have the same support, then the function $A(s) = \int p_1^s p_0^{1-s} \, d\mu$ is analytic and strictly convex, hence a minimizer $\sigma \in [0, 1]$ of $A(s)$ exists, and the infimum is, in fact, a minimum. However, if the supports are different, then $A(s)$ may be discontinuous at the endpoints of the interval $[0, 1]$. Hence, a minimizer need not exist, and the r.h.s. in (8) is only an infimum. The proof of our main theorem, Theorem 2.2 below, uses the above classical result for the general case of possibly different support.

We intend to investigate the asymptotic behavior of the Bayesian error probability in the case where the hypotheses are quantum states on $\mathcal{B}(\mathcal{H})$, where $\dim \mathcal{H} = d < \infty$. In order to derive the optimal asymptotic rate exponent, we replace the target function in the variational formula (3) or (8), which defines the classical Chernoff bound, by

$$\hat{A}(s) := \text{Tr}[\rho_0^{1-s} \rho_1^s], \qquad s \in [0, 1].$$

Our main theorem, formulated below, confirms that the logarithm of the infimum of $\hat{A}(s)$ over $[0, 1]$ gives a lower bound on the optimal quantum error exponent.

THEOREM 2.2 (Quantum Chernoff lower bound). *Let $\rho_0, \rho_1$ be two density operators representing quantum states on a finite-dimensional complex Hilbert space $\mathcal{H}$. Then, any sequence of test projections $\Pi_n \in \mathcal{B}(\mathcal{H}^{\otimes n})$, $n \in \mathbb{N}$, satisfies*

$$(9) \qquad \liminf_{n \to \infty} \frac{1}{n} \log \text{Err}(\Pi_n) \ge \inf_{0 \le s \le 1} \log \text{Tr}[\rho_0^{1-s} \rho_1^s].$$

We point out that, indeed, $\hat{A}(s)$ represents the proper generalization of (4) in the context of symmetric hypothesis testing. As already noted in the Introduction, and as conjectured in [17], it turns out to be achievable (see [1]).

It is of interest to evaluate the quantum Chernoff bound for special cases and to investigate its properties as a distinguishability measure for quantum states. In the classical case, it is well known that if $P_i$ are normal laws $N(\mu_i, \sigma^2)$, then the r.h.s. of (8) is $(\mu_1 - \mu_0)/8\sigma^2$. The discussion of quantum Gaussian states would require admitting an infinite-dimensional complex Hilbert space $\mathcal{H}$, and, thus it is outside the scope of our paper. It can be conjectured, however, that our method of proof readily generalizes to an infinite-dimensional setting. Anticipating such a generalized bound, Calsamiglia et al. [4] have recently written down the appropriate analog of the r.h.s. of (9) for Gaussian states and evaluated it for various examples of Gaussian states of light (cf. also [15]). Another discussion of the geometric properties of the quantum Chernoff bound and a derivation of the related quantum Hoeffding bound can be found in [2].

**3. Proof of the main theorem.** We will prove Theorem 2.2, applying the corresponding classical result, Theorem 2.1, to appropriate probability distributions appearing in the general noncommutative setting.

PROOF OF THEOREM 2.2. We will establish

$$\liminf_{n \to \infty} \frac{1}{n} \log(\mathrm{Err}(\Pi_n)) \geq \inf_{0 \leq s \leq 1} \log \mathrm{Tr}[\rho_0^{1-s} \rho_1^s],$$

for any sequence of projections $\Pi_n \in \mathcal{B}(\mathcal{H}^{\otimes n})$, $n \in \mathbb{N}$.

We consider two arbitrary density operators $\rho_0, \rho_1$ on a finite-dimensional Hilbert space $\mathcal{H} = \mathbb{C}^d$ with spectral representations

$$\rho_0 = \sum_{i=1}^d \lambda_i |x_i\rangle\langle x_i|, \qquad \rho_1 = \sum_{i=1}^d \gamma_i |y_i\rangle\langle y_i|,$$

that is, $|x_i\rangle$, $i = 1, \ldots, d$, and $|y_i\rangle$, $i = 1, \ldots, d$, are two orthonormal bases (ONB) of eigenvectors in $\mathbb{C}^d$, and $\lambda_i, \gamma_i \in [0, 1]$ are the respective eigenvalues of $\rho_0$ and $\rho_1$.

Let $\Pi$ be a projection onto a subspace of $\mathbb{C}^d$, then

$$\mathrm{Tr}[\Pi \rho] = \mathrm{Tr}\left[\Pi \left(\sum_{i=1}^d \lambda_i |x_i\rangle\langle x_i|\right)\right] = \sum_{i=1}^d \lambda_i \langle x_i | \Pi x_i\rangle$$

$$= \sum_{i=1}^d \lambda_i \|\Pi x_i\|^2 = \sum_{i=1}^d \lambda_i \sum_{j=1}^d |\langle \Pi x_i | y_j\rangle|^2,$$

where the third identity is true since $\Pi$ is a projection, and the last one is by Parseval's identity for the ONB $|y_j\rangle$, $j = 1, \ldots, d$. In the same way, we obtain

$$\mathrm{Tr}[(\mathbf{1} - \Pi)\rho_1] = \sum_{j=1}^{d} \gamma_j \sum_{i=1}^{d} |\langle(\mathbf{1} - \Pi)y_j|x_i\rangle|^2.$$

Now, in view of the identity $|\langle(\mathbf{1} - \Pi)y_j|x_i\rangle|^2 = |\langle(\mathbf{1} - \Pi)x_i|y_j\rangle|^2$, we have

$$\mathrm{Err}(\Pi) = \tfrac{1}{2}(\mathrm{Tr}[\rho_0\Pi] + \mathrm{Tr}[\rho_1(\mathbf{1} - \Pi)])$$

$$= \tfrac{1}{2} \sum_{i,j=1}^{d} (\lambda_i |\langle \Pi x_i|y_j\rangle|^2 + \gamma_j |\langle(\mathbf{1} - \Pi)x_i|y_j\rangle|^2).$$

Denote $a = \langle \Pi x_i|y_j\rangle$ and $b = \langle(\mathbf{1} - \Pi)x_i|y_j\rangle$. Since for any complex $a, b$ the inequality $|a|^2 + |b|^2 \geq |a + b|^2/2$ holds, we obtain from the last display

$$(10) \qquad \mathrm{Err}(\Pi) \geq \sum_{i,j=1}^{d} \tfrac{1}{4} \min\{\lambda_i, \gamma_j\} |\langle x_i|y_j\rangle|^2.$$

Note that

$$(11) \qquad p_{i,j} := \lambda_i |\langle x_i|y_j\rangle|^2, \qquad q_{i,j} := \gamma_j |\langle x_i|y_j\rangle|^2, \qquad i,j = 1, \ldots, d,$$

define probability measures $P$ and $Q$ on $d^2$ elements, respectively. Indeed,

$$\sum_{i,j=1}^{d} p_{i,j} = \sum_{i,j=1}^{d} \lambda_i |\langle x_i|y_j\rangle|^2 = \sum_{i=1}^{d} \lambda_i \|x_i\|^2 = \sum_{i=1}^{d} \lambda_i = 1,$$

and similarly for $(q_{i,j})$. Now, inequality (10) may be written

$$(12) \qquad \mathrm{Err}(\Pi) \geq \tfrac{1}{4} \sum_{i,j=1}^{d} \min\{p_{i,j}, q_{i.j}\}.$$

Observe, according to (6) and (7), the r.h.s. above is up to the factor $1/2$ equal to the classical minimal Bayesian error probability $\Delta(P, Q)$ of discrimination between probability measures $P$ and $Q$:

$$(13) \qquad \tfrac{1}{2} \sum_{i,j=1}^{d} \min\{p_{i,j}, q_{i,j}\} = \Delta(P, Q).$$

Next, we consider the case where the quantum hypotheses are $\rho_0^{\otimes n}$ and $\rho_1^{\otimes n}$. Then, the corresponding classical probability measures according to (11) are product measures $P^n$ and $Q^n$, for $P, Q$ corresponding to $\rho_0, \rho_1$, respectively. Applying inequality (12), (13) and subsequently combining it with the classical result on the Chernoff bound for $\Delta(P^n, Q^n)$, Theorem 2.1, we obtain for any sequence of projections $\Pi_n \in \mathcal{B}(\mathcal{H}^{\otimes n})$, $n \in \mathbb{N}$,

$$\liminf_{n \to \infty} \frac{1}{n} \log \mathrm{Err}(\Pi_n) \geq \lim_{n \to \infty} \frac{1}{n} \log\left(\frac{1}{2}\Delta(P^n, Q^n)\right)$$

$$= \log\left(\inf_{0 \leq s \leq 1} \sum_{i,j=1}^{d} p_{i,j}^{1-s} q_{i,j}^{s}\right).$$

We finish the proof by verifying

$$\sum_{i,j=1}^{d} p_{i,j}^{1-s} q_{i,j}^{s} = \sum_{i,j=1}^{d} \lambda_i^{1-s} \gamma_j^s |\langle x_i | y_j \rangle|^2 = \sum_{i,j=1}^{d} \lambda_i^{1-s} \langle x_i | y_j \rangle \gamma_j^s \langle y_j | x_i \rangle$$

$$= \mathrm{Tr}\left[\sum_{i,j=1}^{d} \lambda_i^{1-s} |x_i\rangle\langle x_i| \; \gamma_j^s |y_j\rangle\langle y_j|\right] = \mathrm{Tr}[\rho_0^{1-s} \rho_1^s]. \qquad \square$$

## APPENDIX

As announced in Section 2, we give a proof for Theorem 2.1 for the general case where the two probability measures involved are allowed to have different supports. As far as possible, we follow the proof in the case of same support by Chentsov and Morozova [5].

PROOF OF THEOREM 2.1. 1. *Preliminary observations.* Assume that two probability measures $P_0$, $P_1$ on a measurable space $(\Omega, \Sigma)$ have support $D_i = \mathrm{supp}(P_i)$, $i = 0, 1$. Denote $B = D_1 \cap D_2$, and for $i = 0, 1$,

(14) $$S_i = D_i \setminus B.$$

We introduce the measure $\mu = P_0 + P_1$ and define the densities $p_i = dP_i/d\mu$, $i = 0, 1$. Then, clearly $p_1 + p_2 = 1$. We assume the densities and the sets $D_i$ are chosen such that

$$D_i = \omega : p_i(\omega) > 0, \qquad i = 0, 1,$$

hence

$$B = \omega : p_0(\omega) > 0, p_1(\omega) > 0.$$

Recall the definition of the Hellinger arc of densities for parameter $s \in [0, 1]$:

$$p_s(\omega) = p_1^s(\omega) p_0^{1-s}(\omega) A^{-1}(s),$$

where

$$A(s) = \int p_1^s(\omega) p_0^{1-s}(\omega) \mu(d\omega)$$

is a normalizing factor. Note that for $s=0$ and $s=1$, we obtain the initial densities $p_0$, $p_1$ respectively, so that $A(0) = A(1) = 1$. However, the function $A(s)$ is not continuous in general at the endpoints $0, 1$. Indeed, the integral is over the set $B$,

$$A(s) = \int_B p_1^s(\omega) p_0^{1-s}(\omega) \mu(d\omega),$$

and by dominated convergence it follows that

$$A_+(0) := \lim_{s \searrow 0} A(s) = \int_B p_0(\omega) \mu(d\omega) = P_0(B),$$

$$A_-(1) := \lim_{s \nearrow 1} A(s) = \int_B p_1(\omega) \mu(d\omega) = P_1(B).$$

Furthermore, observe that for $s \in (0,1)$ the densities $p_s$ have support $B$, with limits at the endpoints

$$p_{0+}(\omega) = p_0(\omega)/P_0(B), \qquad p_{1-}(\omega) = p_1(\omega)/P_1(B).$$

Hence, the corresponding limiting measures are the conditional probability measures

$$P_{0+}(\cdot) = P_0(\cdot|B), \qquad P_{1-}(\cdot) = P_1(\cdot|B).$$

If the sample space is restricted to $B$, the densities $p_s$, $s \in (0,1)$, can be written in exponential family form:

(15) $$p_s(\omega) = \exp\left( s \log \frac{p_1(\omega)}{p_0(\omega)} \right) p_0(\omega) A^{-1}(s), \qquad \omega \in B.$$

For $s = 0, 1$, the above holds if $B = D_s$. Also, for $s = 0, 1$, if $B \neq D_s$, then the restriction $p_s|B$ is not a probability density. We denote

$$H(s) = \log A(s), \qquad H_+(0) = \log P_0(B), \qquad H_-(1) = \log P_1(B).$$

2. *Bayesian error probabilities* $\mathrm{Err}(\lambda_n^*)$ *by change of measure to* $P_s$. Recall the form of the optimal test $\lambda_n^*$ on $\Omega^n$ for equiprobable hypothetic densities $p_0$ and $p_1$ on $\Omega$:

$$\lambda_n^* = \mathbf{1}\left\{ \prod_{j=1}^n p_1(\omega_j) > \prod_{j=1}^n p_0(\omega_j) \right\},$$

where $\omega_1, \ldots, \omega_n$ are $n$ i.i.d. observations. (One may also take "$\geq$" or decide arbitrarily on the "=" set.) We partition the set $\Omega^n$ into disjoint subsets $S_{0,n}$, $S_{1,n}$ and $B_n$:

$$S_{0,n} := \{\text{there is } j \in \{1, \ldots, n\} \text{ such that } \omega_j \in S_0\},$$
$$S_{1,n} := \{\text{there is } j \in \{1, \ldots, n\} \text{ such that } \omega_j \in S_1\},$$

where $S_i$, $i=0,1$, were defined in (14). The remaining case is the event

$$B_n := \{\omega^n \in \Omega \colon \omega_j \in B \text{ for } j=1,\ldots,n\}.$$

Denote $\omega^n = (\omega_1,\ldots,\omega_n) \in \Omega^n$. We have $\lambda_n^*(\omega^n) = 1$ (decision in favor of $P_1$) if $\omega^n \in S_{1,n}$, that is, an event happens which excludes $P_0$. Similarly, we have $\lambda_n^*(\omega^n) = 0$ for $\omega^n \in S_{0,n}$. For $\omega^n \in B_n$, define the (normed) log-likelihood ratio by

$$L_n(\omega^n) := n^{-1} \sum_{i=1}^n \log \frac{p_1}{p_0}(\omega_i).$$

Then, we can describe the test $\lambda_n^*$

(16) $$\lambda_n^*(\omega^n) = \mathbf{1}\{L_n(\omega^n) > 0, \omega^n \in B_n\} + \mathbf{1}\{\omega^n \in S_{1,n}\}.$$

Further, we define, for $i=0,1$, functions

$$G_{s,n}^{(i)}(\omega^n) = \mathbf{1}\{\omega^n \in B_n\} n^{-1} \sum_{j=1}^n \log \frac{p_i}{p_s}(\omega_j).$$

We note the following relations, for $\omega \in B$:

(17) $$\log \frac{p_0}{p_s}(\omega) = -s \log \frac{p_1}{p_0}(\omega) + H(s),$$

(18) $$\log \frac{p_1}{p_s}(\omega) = (1-s) \log \frac{p_1}{p_0}(\omega) + H(s).$$

To prove (18), observe that

$$\log \frac{p_1}{p_s} = \log \frac{p_1 A(s)}{\exp(s \log p_1/p_0) p_0} = \log \frac{p_1}{p_0} - s \log \frac{p_1}{p_0} + H(s)$$
$$= (1-s) \log \frac{p_1}{p_0} + H(s).$$

Furthermore, it holds

$$\log \frac{p_0}{p_s} = \log \frac{p_0 A(s)}{\exp(s \log p_1/p_0) p_0} = -s \log \frac{p_1}{p_0} + H(s),$$

which implies (17). As a consequence of (17) and (18), we have, for $\omega^n \in B_n$,

(19) $$G_{s,n}^{(0)}(\omega^n) = -s L_n(\omega^n) + H(s),$$

(20) $$G_{s,n}^{(1)}(\omega^n) = (1-s) L_n(\omega^n) + H(s).$$

In the sequel, we write $E_s$ for expectation under the density $p_s$ and denote by $E_s^n$ the expectation under the product density for the respective basic density $p_s$. Notice that the test $\lambda_n^*$ necessarily decides correctly if $\omega^n \in B_n^c =$ $S_{0,n} \cup S_{1,n}$. Thus, the minimal Bayesian error probabilities can be expressed, for any $s \in (0,1)$, as

$$\text{Err}(\lambda_n^*) = E_0^n \lambda_n^* + E_1^n(1-\lambda_n^*) = E_0^n \mathbf{1}_{B_n} \lambda_n^* + E_1^n \mathbf{1}_{B_n}(1-\lambda_n^*) \quad (21)$$

$$= E_s^n \lambda_n^* \exp(nG_{s,n}^{(0)}) + E_s^n(1-\lambda_n^*)\exp(nG_{s,n}^{(1)})$$

$$= E_s^n \lambda_n^* \exp(-nsL_n + nH(s))$$

$$+ E_s^n(1-\lambda_n^*)\exp(n(1-s)L_n + nH(s))$$

$$= \exp(nH(s))\{E_s^n(\lambda_n^*\exp(-nsL_n) + (1-\lambda_n^*)\exp(n(1-s)L_n))\}. \quad (22)$$

3. *Upper risk bound.* From the expression (16) for $\lambda_n^*$, we see that, for all $\omega^n \in B_n$,

$$\lambda_n^* \exp(-nsL_n) + (1-\lambda_n^*)\exp(n(1-s)L_n) \leq 1,$$

so that (22) implies, for all $n \in \mathbb{N}$,

$$\text{Err}(\lambda_n^*) \leq \exp(nH(s))$$

and, hence,

$$\frac{1}{n}\log \text{Err}(\lambda_n^*) \leq H(s).$$

Since $s \in (0,1)$ was arbitrary, and since the bounds $H(0) = H(1) = 0$ are trivial, we obtain

$$\frac{1}{n}\log \text{Err}(\lambda_n^*) \leq \inf_{0 \leq s \leq 1} H(s).$$

4. *Convexity of $H(s)$ on $(0,1)$.* Using the exponential family expression (15) for densities $p_s$, the function $H(s)$ may be written for $s \in (0,1)$,

$$H(s) = \log \int_B \exp\left(s \log \frac{p_1(\omega)}{p_0(\omega)}\right) p_0(\omega)\, d\mu(\omega). \quad (23)$$

It follows

$$H'(s) = \frac{A'(s)}{A(s)} = \frac{\int_B \log p_1(\omega)/p_0(\omega) \exp(s \log p_1(\omega)/p_0(\omega)) p_0(\omega)\, d\mu(\omega)}{A(s)},$$

where the fact that $A(s)$ can be differentiated under the integral sign, and the integral is finite for all $s \in (0,1)$, is from the basic theory of exponential families. In the sequel, we identify expectation under $p_s$ and its restriction $p_s|B$ for $s \in (0,1)$. We can thus write (for a random variable $\omega$ taking values in $B$)

$$H'(s) = E_s \log \frac{p_1(\omega)}{p_0(\omega)} = E_s \log \frac{p_s(\omega)}{p_0(\omega)} - E_s \log \frac{p_s(\omega)}{p_1(\omega)}. \quad (24)$$

For the second derivative, we obtain

$$H''(s) = \frac{A''(s)A(s) - (A'(s))^2}{A^2(s)}$$

$$= \frac{\int (\log p_1(\omega)/p_0(\omega))^2 \exp(s \log p_1(\omega)/p_0(\omega)) p_0(\omega) \, d\mu(\omega)}{A(s)} - (H'(s))^2$$

$$= E_s \left( \log \frac{p_1(\omega)}{p_0(\omega)} \right)^2 - \left( E_s \log \frac{p_1(\omega)}{p_0(\omega)} \right)^2 \geq 0,$$

since the last expression is the variance of the random variable $\log(p_1/p_0)(\omega)$ under $p_s$. Thus, $H(s)$ is convex on $(0, 1)$. There are two cases.

**Case 1.** There is some $s \in (0, 1)$ such that $H''(s) = 0$. Then, $\log(p_1/p_0)(\omega)$ is constant $P_s$-almost surely. Since all $P_s$, $s \in (0, 1)$, dominate each other, $(p_1/p_0)(\omega)$ is also constant $P_s$-almost surely, for all $s \in (0, 1)$ and $H''(s) = 0$ for all these $s$. Hence, $H(s)$ is linear on $(0, 1)$. Furthermore, each $P_s$, $s \in (0, 1)$, dominates $\mu$ on $B$ (i.e., dominates $\mu|B$). It follows

$$\frac{p_1}{p_0}(\omega) = c, \qquad \mu\text{-a.s. on } B,$$

for some constant $c > 0$. In that case,

$$P_1(B) = \int_B c \, dP_0 = c P_0(B)$$

and

$$c = \frac{P_1(B)}{P_0(B)}.$$

This implies

$$P_0(\cdot|B) = P_1(\cdot|B) = P_s, \qquad s \in (0, 1),$$

(25) $$A(s) = (P_0(B))^{1-s}(P_1(B))^s, \qquad s \in (0, 1).$$

**Case 2.** For all $s \in (0, 1)$, we have $H''(s) > 0$. Then, $H(s)$ is strictly convex on $(0, 1)$.

5. *Lower risk bound.* Since, according to (24), for arbitrary $s \in (0, 1)$,

$$H'(s) = E_s \log \frac{p_1}{p_0}(\omega),$$

we have in view of (19) and (20), for each $n \in \mathbb{N}$,

$$E_s^n G_{s,n}^{(0)} = -sH'(s) + H(s) =: \gamma_0(s),$$

$$E_s^n G_{s,n}^{(1)} = (1-s)H'(s) + H(s) =: \gamma_1(s).$$

Since $G_{s,n}^{(i)}$ is an i.i.d. average, we have by the Law of Large Numbers, as $n$ tends to infinity,

$$G_{s,n}^{(0)}(\omega^n) \to \gamma_0(s), \qquad G_{s,n}^{(1)}(\omega^n) \to \gamma_1(s),$$

almost surely under $P_s$. Let $\delta, \eta > 0$ be arbitrary and consider the subsets

$$U_n := \{\omega^n : G_{s,n}^{(i)}(\omega^n) - \gamma_i(s) \geq -\eta, i = 0, 1\}, \qquad n \in \mathbb{N}.$$

Then, again by the Law of Large Numbers, there is an $n_\delta \in \mathbb{N}$ such that

$$P_s^n(U_n) \geq 1 - \delta \qquad \text{for all } n \geq n_\delta.$$

Starting with identity (21), we estimate the minimal error probability for $n \geq n_\delta$:

$$\begin{aligned}
\operatorname{Err}(\lambda_n^*) &= E_s^n \lambda_n^* \exp(nG_{s,n}^{(0)}) + E_s^n(1 - \lambda_n^*) \exp(nG_{s,n}^{(1)}) \\
&\geq E_s^n \mathbf{1}\{U_n\}(\lambda_n^* \exp(n\gamma_0(s) - n\eta) + (1 - \lambda_n^*)\exp(n\gamma_1(s) - n\eta)) \\
&\geq E_s^n \mathbf{1}\{U_n\} \exp(n\min(\gamma_0(s), \gamma_1(s)) - n\eta) \\
&\geq (1 - \delta) \exp(n\min(\gamma_0(s), \gamma_1(s)) - n\eta).
\end{aligned}$$

Consequently, we have, for any sequence of test functions $\lambda_n$, $n \in \mathbb{N}$,

$$\liminf_{n \to \infty} n^{-1} \log \operatorname{Err}(\lambda_n) \geq \min(\gamma_0(s), \gamma_1(s)) - \eta.$$

Since $\eta$ was arbitrary, we obtain for any $s \in (0, 1)$

$$\liminf_{n \to \infty} n^{-1} \log \operatorname{Err}(\lambda_n) \geq \min(\gamma_0(s), \gamma_1(s)),$$

and, hence,

$$\liminf_{n \to \infty} n^{-1} \log \operatorname{Err}(\lambda_n) \geq \sup_{0 < s < 1} \min(\gamma_0(s), \gamma_1(s)).$$

It remains to show that

(26) $$\sup_{0 < s < 1} \min(\gamma_0(s), \gamma_1(s)) \geq \inf_{0 \leq s \leq 1} H(s).$$

Recall that the values $H'(s)$ are well defined for $s \in (0, 1)$ and that $H(s)$ is convex in that domain. Hence, there exist limits

$$H'_+(0) = \lim_{s \searrow 0} H'(s), \qquad H'_-(1) = \lim_{s \nearrow 1} H'(s).$$

Observe that the limits are possibly infinite. However, due to convexity, only $H'_+(0) = -\infty$ or $H'_+(1) = \infty$ may occur.

Again, in view of the convexity of $H(s)$ on $(0, 1)$, the following cases may occur:

(a) $H'_+(0) < 0$, $H'_-(1) > 0$,

(b) $H'_+(0) < 0$, $H'_-(1) \leq 0$,
(c) $H'_+(0) \geq 0$, $H'_-(1) > 0$,
(d) $H'_+(0) \geq 0$, $H'_-(1) \leq 0$.

**Case (a).** In this case, $H$ cannot be linear, so that due to the above discussion in 4 (involving Cases 1 and 2) it is strictly convex in $(0,1)$. Hence, there is a unique minimum of $H$ on $[0,1]$ at some $\sigma \in (0,1)$ with $H'(\sigma) = 0$. We have

$$\gamma_0(\sigma) = \gamma_1(\sigma) = H(\sigma),$$

hence,

$$\sup_{0 < s < 1} \min(\gamma_0(s), \gamma_1(s)) \geq H(\sigma) = \inf_{0 \leq s \leq 1} H(s).$$

**Case (b).** Again, due to convexity, the infimum of $H$ on $[0,1]$ is attained (uniquely) at $s \nearrow 1$:

$$\inf_{0 \leq s \leq 1} H(s) = \lim_{s \nearrow 1} H(s) = H_-(1).$$

Now, for $s \in (0,1)$ we have $H'(s) \leq 0$, and, hence,

$$\gamma_0(s) = -sH'(s) + H(s) \geq H(s) \geq (1-s)H'(s) + H(s) = \gamma_1(s),$$

which implies

$$\sup_{0 < s < 1} \min(\gamma_0(s), \gamma_1(s)) \geq \sup_{0 < s < 1} \gamma_1(s) \geq \limsup_{s \nearrow 1} \gamma_1(s)$$

$$\geq H_-(1) = \inf_{0 \leq s \leq 1} H(s).$$

**Case (c).** This is symmetric to case (b). We obtain

$$\inf_{0 \leq s \leq 1} H(s) = H_+(0)$$

and

$$\sup_{0 < s < 1} \min(\gamma_0(s), \gamma_1(s)) \geq H_+(0) = \inf_{0 \leq s \leq 1} H(s).$$

Now, for $s \in (0,1)$, we have $H'(s) \geq 0$, and, hence,

$$\gamma_1(s) = (1-s)H'(s) + H(s) \geq H(s) \geq -sH'(s) + H(s) = \gamma_0(s)$$

which implies

$$\sup_{0 < s < 1} \min(\gamma_0(s), \gamma_1(s)) \geq \sup_{0 < s < 1} \gamma_0(s) \geq \limsup_{s \searrow 0} \gamma_0(s)$$

$$\geq H_+(0) = \inf_{0 \leq s \leq 1} H(s).$$

**Case (d).** Due to convexity, we must have $H'_+(0) = H'_-(1) = 0$; then, $H(s)$ is constant on $(0,1)$. By (25), we then have $P_0(B) = P_1(B)$ and

$$H(s) = \log P_0(B) = \log P_1(B), \qquad s \in (0,1).$$

Consequently,

$$\gamma_0(s) = \gamma_1(s) = H(s) = \inf_{0 \leq s \leq 1} H(s),$$

and we obtain trivially

$$\sup_{0 < s < 1} \min(\gamma_0(s), \gamma_1(s)) \geq \inf_{0 \leq s \leq 1} H(s).$$

We have verified inequality (26) in all cases (a)–(d). Hence, for any sequence of test functions $\lambda_n$ on $\Omega^n$, $n \in \mathbb{N}$, we have

$$\liminf_{n \to \infty} n^{-1} \log \operatorname{Err}(\lambda_n) \geq \liminf_{n \to \infty} n^{-1} \log \operatorname{Err}(\lambda_n^*) \geq \inf_{0 \leq s \leq 1} H(s).$$

The upper and lower bounds together complete the proof. □

**Acknowledgments.** The first author wishes to thank Ruedi Seiler and the Mathematical Physics group of Technical University Berlin for the opportunity to spend a research semester there. Harrison Zhou contributed a shortening of the main proof.

The second author is grateful to Ruedi Seiler and her colleagues Nihat Ay, Rainer Siegmund-Schultze, Markus Müller and Tyll Krüger for any kind of support. Further, she wants to thank Profs. I. Csiszár and Dénes Petz. The idea to elaborate on the topic of the present paper goes back to stimulating discussions in the Information Theory Seminar at the Rényi Institute, Budapest.

DEPARTMENT OF MATHEMATICS
CORNELL UNIVERSITY
MALOTT HALL
ITHACA, NEW YORK 14853
USA
E-MAIL: nussbaum@math.cornell.edu

MAX PLANCK INSTITUTE FOR MATHEMATICS
IN THE SCIENCES
INSELSTRASSE 22
04103 LEIPZIG
GERMANY
E-MAIL: szkola@mis.mpg.de